\documentclass[useAMS,usenatbib]{mn2e}

\usepackage[usenames,dvipsnames]{xcolor}

\usepackage{aas_macros}
\usepackage{amssymb,amsmath}
\usepackage{epsf}
\usepackage{bm}
\usepackage{natbib}
\usepackage{graphicx}
\usepackage{cancel}
\usepackage{ulem}

\title[New class of neutron stars]
{New possible class of neutron stars: hot and fast
non-accreting rotators}

\author[A. I. Chugunov, M. E. Gusakov, E.M. Kantor]
{A.~I.~Chugunov$^1$\thanks{andr.astro@mail.ioffe.ru},
 M. E. Gusakov$^{1,2}$\thanks{gusakov@astro.ioffe.ru}
 E. M. Kantor$^1$\thanks{kantor@mail.ioffe.ru}\\
 $^1$Ioffe Institute, St Petersburg, Russia\\
 $^2$St.-Petersburg State Polytechnical University,
Polytekhnicheskaya 29, 195251 St.-Petersburg, Russia }

\begin{document}

\date{Accepted 2014 xxxx. Received 2014 xxxx;
in original form 2014 xxxx}

\pagerange{\pageref{firstpage}--\pageref{lastpage}}
\pubyear{2014}

\maketitle

\label{firstpage}

\begin{abstract}
A new class of neutron stars (NSs) -- hot rapidly rotating
non-accreting NSs, which we propose to call HOFNARs (HOt
and Fast Non-Accreting Rotators) or ``\textit{hot
widows}''\ (in analogy with ``black widow'' pulsars) -- is
suggested. We argue that such stars should originate from
the low-mass X-ray binaries (LMXBs) provided that they were
unstable with respect to excitation of $r$-modes at the end
of accretion epoch (when their low-mass companions ceased
to fill the Roche lobe). High temperature of ``hot
widows''/HOFNARs is maintained by $r$-mode dissipation
rather than by accretion. We analyse observational
properties of ``hot widows''/HOFNARs and demonstrate that
these objects form a specific separate class of neutron
stars. In particular, some of the most stable X-ray sources
among the candidates to quiescent LMXB systems (qLMXBs),
can, in fact, belong to that new class. We formulate
observational criteria which allow to distinguish ``hot
widows''/HOFNARs from qLMXB systems, and argue that
available observations of X-ray sources 47 Tuc X5 and X7
satisfy (or, at least, do not contradict) these criteria.
In addition, we discuss indirect evidences in favor of
``hot widows''/HOFNARs existence, following from the
analysis of observations and predictions of population
synthesis theories. If that new class of NSs does exist, it
would prove the possibility to emit gravitational waves by
mass-current multipole. Various applications of our
results, such as prospects for constraining superdense
matter properties with ``hot widows''/HOFNARs, are
analyzed.
\end{abstract}


\section{Introduction} \label{Sec_Introd}

Neutron stars (NSs) are usually divided into various
classes according to their diverse observational
manifestations (e.g., \citealt{Harding13}),
although, to all appearances, the equation of state in their
interiors is universal
(e.g., \citealt*{hpy07}). Furthermore,
it is generally believed that,
depending on the evolution stage,
the same star
can belong to different classes
(e.g.\ \citealt{Vigano_etal13_Unif,Papitto_etal13}).
Detailed
observations of NSs from
each
class together with their accurate theoretical modeling may
shed light on still poorly known
properties of superdense matter.
For example,
it is very hard to explain pulsar glitches without
invoking
nucleon superfluidity in their interiors
(e.g., \citealt{pa85}),
while
comparison of the theory with observations
of cooling isolated NSs
allowed us to put tight
constraints
on
the parameters of
nucleon superfluidity
in the NS core (e.g., \citealt{gkyg04,page04}).
These constraints have been subsequently confirmed by recent
observations of a real-time cooling NS
in Cassiopeia A supernova remnant
(\citealt{page11,shternin11}).%
\footnote{Observations of this NS have been recently
reanalyzed by \citet{ppsk13}, who argued that its cooling
may not be so strong, as it was reported previously by
\citet{hh10} and \citet{shternin11}. If these new results
are correct, parameters of the superfluid model of
\citet{shternin11} should be slightly modified, which does
not exclude superfluidity \textit{per se} (for example,
similar model of \citet{gkyg04} predicts slower cooling for
this NS).}
As shown by \cite{Kantor11} and \cite{ha12},
the dependence of superfluid properties of matter on temperature
can reveal itself in
a deviation of the pulsar braking index
from the standard value 3,
predicted by the
magneto-dipole model.
Binary systems with pulsars
serve as the excellent targets for
reliable measurements
of NS masses (see, e.g.,
\citealt{Demorest_etal10,Antoniadis_etal13}), while
observations of hot NSs in LMXBs
can be used to estimate both mass
and radius for these stars
(see, e.g.,
\citealt{Guillot_etal13})
and, possibly, can reveal
resonance features in their oscillation spectra
(\citealt*{gck14_short,gck13_large}).

\begin{figure*}
    \begin{center}
        \leavevmode
        \epsfxsize=18cm \epsfbox{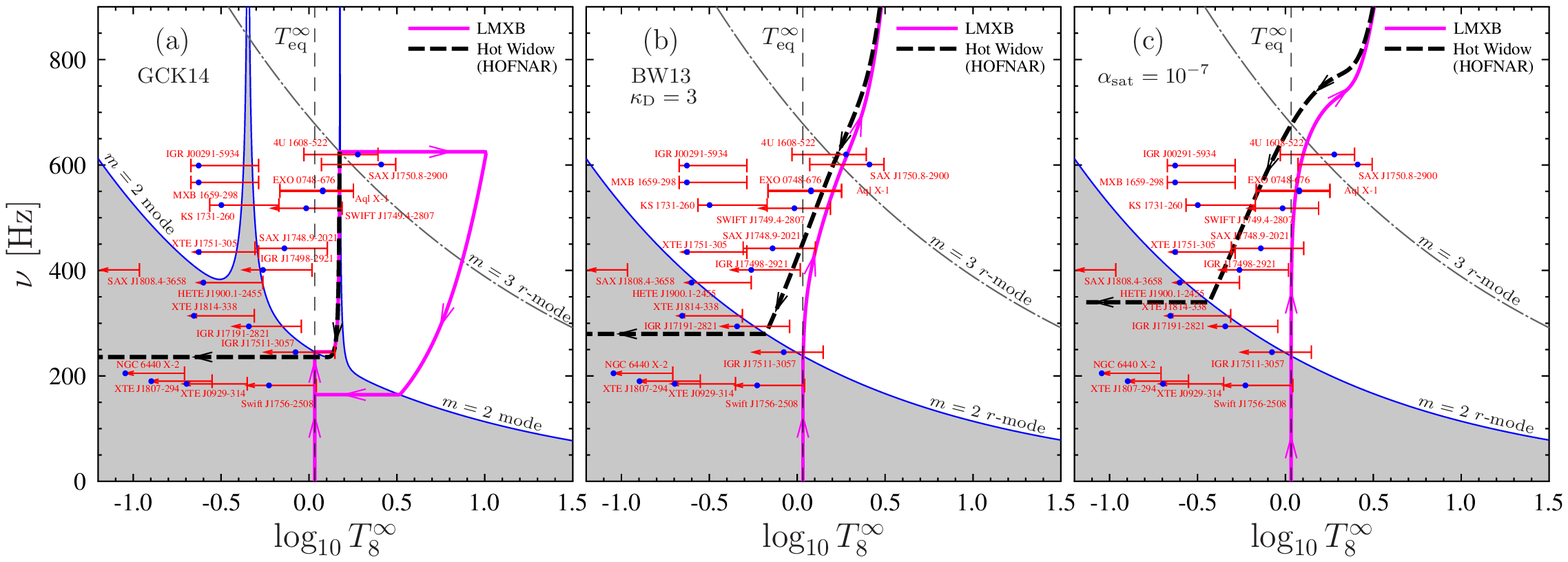}
    \end{center}
    \caption{Evolution of frequency $\nu$ and internal redshifted  temperature $T^\infty$
    for an accreting NS in LMXB
    (thick solid line)
    and for non-accreting NS with exhausted low-mass companion
    (thick dashed line),
    for three models of $r$-mode instability:
    (a) model of resonant interaction of superfluid inertial modes with $r$-mode
    (\citealt{gck14_short,gck13_large}, GCK14);
    (b) model of \citealt{bw13} (BW13)
    with nonlinear mode coupling parameter
    $\kappa_\mathrm D=3$;
    (c) widely used model in which the $r$-mode saturation amplitude is constant
    and is taken to be $\alpha_\mathrm{sat}=10^{-7}$.
    Arrows in the curves demonstrate direction of the NS evolution.
    Thin solid lines show instability curves for the
    most unstable (at a given temperature) mode with $m=2$;
    the dot-dashed line
    demonstrates instability curve for $m=3$ $r$-mode.
    The stability region is shaded in grey;
    in the white region at least one of the oscillation modes is unstable.
    Vertical dashed line indicates $T^\infty_\mathrm{eq}$.
    Temperatures and frequencies
    of the NSs observed in LMXBs are shown by filled circles,
    error bars show uncertainties due to unconstrained NS envelope
    composition (see\ \citealt{gck14_short}  and references
    therein).}
    \label{Fig_Evol}
\end{figure*}

In this paper we consider a hypothesis of existence of hot
(effective redshifted surface temperatures
$T^\infty_\mathrm{eff}\sim 10^6$~K, internal redshifted
temperatures $T^\infty\sim 10^8$~K) nonaccreting rapidly
rotating (spin frequencies $\nu\gtrsim 200$~Hz) NSs, whose
temperature stays high during $\sim 10^9$~yrs because of
the  $r$-mode instability. Such stars can be descendants of
NSs in LMXBs (see, e.g.\ \citealt{rb03,bw13,gck14_short}
and Sec.\ \ref{Sec_Evol}) and, as we will show here, they
form a new class of NSs with very specific physical
properties and observational features (see Sec.\
\ref{Sec_Nab}). Following \cite{gck14_short} we suggest to
call these NSs ``HOFNARs'' (from HOt and Fast Non-Accreting
Rotators) or ``hot widows'' (similar to ``black widow''
pulsars).

High internal temperature $T^\infty\sim 10^8$~K
of ``hot widows''/HOFNARs,
and hence their high luminosity
$L_\mathrm{cool}\sim 10^{34}$~erg\,s$^{-1}$
(in the form of neutrino and thermal radiation from the atmosphere)
during $\tau\sim 10^9$~yr,
requires a lot of energy,
$E\sim L_\mathrm{cool} t\sim 10^{51}$~erg.
In a nonaccreting NS
only rotation can
afford
such a huge energy budget.
In addition,
rotation energy
should be deposited into thermal energy with very
high efficiency ($\sim 30\%$),
which is
achievable
with $r$-mode instability
[see Eq.\ (\ref{dErot_dt}) below],
but hardly possible for other mechanisms.
As a result,
existence of ``hot widows''/HOFNARs is impossible without
unstable $r$-modes.

An instability of NSs with respect to excitation of
$r$-modes was discovered less than twenty years ago by
\cite{andersson98} and \cite{fm98}. It is a particular case
of the instability analysed by \cite{chandrasekhar70a} and
\cite{fs78a,fs78b} [Chandrasekhar-Friedman-Schutz (CFS)
instability]; it is related to the fact that excitation of
$r$-mode decreases stellar angular momentum and rotation
energy (\citealt{fs78a, fs78b,lu01,hl00}). As a result,
gravitational waves emitted by $r$-modes excite (rather
than damp) $r$-modes at {\it arbitrary} spin frequency
$\nu$ (\citealt*{andersson98,fm98,lom98}). An account for
dissipation suppresses instability at low $\nu$. However,
at high enough temperature and spin frequency an NS still
stays unstable (see, e.g., \citealt{lom98,ak01};
\citealt*{hah11,hdh12}; \citealt{Andersson_etal13}). A
corresponding region of temperatures and frequencies is
known as the ``instability window''.
For NSs inside the
instability window  $r$-modes excite spontaneously and
their dissipation becomes an important source of stellar
heating (\citealt{levin99,heyl02}; 
\citealt{gck14_short,gck13_large};
\citealt*{hga14}
).
It allows NSs to stay hot (i.e., to become ``hot
widows''/HOFNARs) even after the end of accretion epoch,
until their spin frequencies become sufficiently small to
make them stable with respect to excitation of $r$-modes.
We discuss evolution and properties of such ``hot
widows''/HOFNARs, as well as their observational
manifestations.

\section{Scenarios of the NS evolution
resulting in ``hot widow''/HOFNAR formation}
\label{Sec_Evol}

In this section we will demonstrate that
``hot widows''/HOFNARs
appear
in all scenarios of NS evolution in LMXBs,
in which the $r$-mode instability
plays essential role
(that is, the probability to observe
an NS with excited $r$-mode
is not small).
For this purpose we will analyse
three models of $r$-mode instability:
($i$) model (a), whose characteristic feature is the
resonance interaction of superfluid inertial modes with the
$r$-mode (for more details see\
\citealt{gck14_short,gck13_large});
 ($ii$) model (b), for
which the saturation of $r$-mode
is defined by the first
parametric instability due to non-linear interaction of
$r$-mode with the higher-order (principal mode number
$n\sim100$) inertial modes, freely penetrating the crust
\citep{bw13};
($iii$) widely used model (c), for which the saturation
amplitude of $r$-modes
(i.e., the maximum achievable $r$-mode amplitude)
is independent of the oscillation frequency and
temperature.

NS evolution at the stage of active accretion in an LMXB is
driven by the equations formulated by
\cite{olcsva98,hl00,gck13_large}.
For the models listed
above the evolution was studied in detail by
\cite{gck14_short,gck13_large} [model (a)],
by \cite{bw13} [model (b)],
by \cite{levin99}, \cite{heyl02}, and
\cite{gck13_large} [model (c)], and is briefly described
here for completeness.
Corresponding evolutionary tracks in the plane
``rotation frequency $\nu$
--- red-shifted internal temperature $T^\infty\,$''
are shown by thick solid lines in Fig.~\ref{Fig_Evol}
(evolution direction is indicated by arrows).
Fig.~\ref{Fig_Evol} also shows boundaries of the stability region
for the most unstable oscillation mode%
\footnote{For models (b) and (c) the most unstable mode, at
any temperature, is the quadrupolar ($m=2$) $r$-mode. Model
(a) accounts for the avoided-crossings between the
superfluid and normal modes which means that, strictly
speaking, different modes can be the most unstable ones at
different temperatures. However, it can be shown that far
from the resonances (stability peaks) the most unstable
oscillation mode is very similar to $r$-mode of a
non-superfluid NS (for more details see
\citealt{gck13_large,gck14_short}). In what follows, for
brevity, the most unstable mode will be referred to as
$r$-mode for all the models.}
with multipolarity $m=2$ (thin solid line) and $r$-mode
with multipolarity $m=3$ (dot-dashed line).
These curves
are calculated in \cite{gck14_short,gck13_large} for a
canonical NS with the mass $M=1.4M_\odot$ and radius
$R=10$~km. Filled circles with error bars in Fig.\
\ref{Fig_Evol} correspond to the observed spin frequencies
and temperatures of NSs (see \citealt{gck14_short} and
references therein).

In the region where all oscillation modes are stable
(filled grey in the figure),
a star is spun up by accretion.
Its temperature $T^\infty=T^\infty_\mathrm{eq}$
(see vertical dashes in the figure)
is determined by the balance
of
stellar cooling and internal heating caused by
nuclear transformations in the accreted NS crust
(deep crustal heating model of \citealt*{bbr98}).
For the range of temperatures considered in this paper an
NS cools down mostly due to neutrino emission from its
core, which is calculated by employing a microphysical
model from \cite{gkyg04}. That model allows one to explain
{\it all} observations of cooling isolated NSs within
minimal assumptions on the NS core composition and
properties. In what follows, in our calculations we
consider an NS with the mass $M=1.4M_\odot$. The accretion
rate, averaged over long period of time, involving both
quiescent and active phases, is chosen to be $\dot
M=3\times10^{-10} M_\odot/\mathrm{yr}$. Such $\dot{M}$
agrees with the estimates of accretion rates for the
hottest NSs in LMXBs
(see
\citealt{hjwt07,heinke_et_al_09,gck14_short} and references
therein) and corresponds to the equilibrium temperature
$T^\infty_\mathrm{eq}\approx 1.08\times 10^8$~K, which does
not depend on the $r$-mode instability model. When a star
reaches the instability region, the amplitude of $m=2$
$r$-mode rapidly grows up, and its dissipation becomes a
powerful source of stellar heating. Subsequent stellar
evolution slightly differs for the model (a) and models (b)
and (c).

In model (a) temperature increase is limited by the
stability peak arising from the resonance interaction of
oscillation modes. Having reached the peak a star spins up
due to accretion,
moving along
the left edge of the
stability peak.
The stellar temperature is maintained by
dissipation of $r$-mode, which is marginally stable and
shown to have the required average amplitude
(\citealt{gck14_short}).
Ascending the stability peak
can be terminated
at \textit{equilibrium} rotation frequency
$\nu_\mathrm{eq}$,
at which the spin up accretion torque
balances the joint
torque produced by the
magneto-dipole losses and gravitational wave emission.
If the accretion torque is sufficiently large,
a star reaches the instability region of $m=3$ $r$-mode,
where that mode excites and rapidly heats the star up.
Gravitational wave emission
then is
so strong
that the star spins down
(despite ongoing accretion)
and returns into the stability region,
where all oscillation modes damp out very fast
and the star cools down to the temperature
$T^\infty=T^\infty_\mathrm{eq}$.
Then the evolution cycle repeats.
A typical period of the
cycle is about $10^8$ years
(i.e.,\ less than the typical duration
of intensive accretion in LMXBs, see, e.g., \citealt{ccth13})
and depends on a spin-up rate of an NS in the course of accretion
(\citealt{gck13_large}).
A star spends a major part of the period climbing up the stability peak.
Thus a probability to find it at the peak
at the end of the LMXB stage (when accretion ceases, e.g.,
due to depletion of the low-mass companion) is high.

In models (b) and (c)
NS evolution in an LMXB is a bit different.
Once entering the instability region,
an NS
rapidly reaches the curve where stellar heating (due to
nuclear reactions in the accreted crust and dissipation of
the saturated $r$-mode) is compensated by neutrino cooling.
Then the star evolves along this curve
(\citealt{levin99,heyl02,gck13_large, bw13}).
The observed
NS temperatures
within these models
can only be explained if one assumes
extremely low saturation amplitude $\alpha\lesssim 10^{-7}$
(see, e.g., \citealt{ms13}).
The maximum allowable NS spin frequency
for such models
is limited by the \textit{equilibrium} frequency
$\nu_\mathrm{eq}$ (see, e.g., \citealt{bw13}).

The LMXB stage does not last forever: As a result of its
own evolution and/or the evolution of orbital period of
LMXB, the low-mass companion eventually ceases to fill the
Roche lobe (see \citealt*{Tauris11,tvh06} and references
therein). This leads to an abrupt termination of accretion,
and switching off heating due to nuclear reactions in the
crust. An NS can then cool down rapidly and become a
millisecond pulsar (MSP). However, if a star was at the
left edge of the stability peak [which is quite probable in
model (a), see above] or in the instability window [which
occurs naturally in models (b) and (c) if $\nu_{\rm eq}$ is
not too small], then it continues to be heated by
dissipation of NS oscillations even after the end of
accretion epoch. As a consequence, an NS stays hot (and
rapidly rotating) for a long time, i.e. it becomes a ``hot
widow''/HOFNAR. Subsequent evolution of an NS with the
excited $r$-mode is described by the same model as that
proposed to study NSs at the LMXB stage
(\citealt{olcsva98,hl00,gck13_large}). The corresponding
evolutionary paths are shown in Fig.\ \ref{Fig_Evol} by the
thick dashes. In model (a) the end of accretion epoch does
not lead to reduction of $T^\infty$, because this would
amplify the NS instability [see\ Fig.\ \ref{Fig_Evol} (a)].
Instead, after accretion stops, oscillation amplitude
automatically adjusts itself (increases) so as to keep the
star at the edge of the stability peak by means of heating
exclusively due to $r$-mode dissipation. In case of models
(b) and (c) the saturated $r$-mode cannot further increase
its amplitude so that absence of accretion results in a
reduction of temperature which is then determined by the
balance between the neutrino luminosity and viscous
$r$-mode heating [see\ Fig.\ \ref{Fig_Evol} (b,c)].

The spin frequency $\nu$
in the beginning of the "hot widow"/HOFNAR phase
is determined by the frequency $\nu$
at the end of LMXB stage%
\footnote{Note that,
in the final stage
of LMXB evolution
an NS can accrete matter being in the so called propeller
regime (\citealt{is75}).
As shown by \cite{Tauris12}, this can lead to
a rapid spin-down of an NS and to corresponding
decreasing of its rotation energy by a factor
of two (or even more).
This effect can reduce amount of NSs that stay unstable
(and thus become ``hot widows''/HOFNARs) after the end of LMXB stage.
};
during that phase an NS simply
slows down
due to
magneto-dipole losses
and
gravitational wave emission
by excited oscillation modes.
Along the evolutionary track
the following condition is satisfied,
\begin{equation}
\left|\tau_\mathrm{GR}\right|=\left|\tau_\mathrm{Diss}\right|,
\label{tau}
\end{equation}
where $\tau_\mathrm{GR}$ and $\tau_\mathrm{Diss}$ are
gravitational-wave and damping timescales, respectively. In
the case of model (a) this condition is fulfilled because
of marginal stability of $r$-mode (see
\citealt{gck13_large}), while for the models (b) and (c) it
determines a ``saturation'' of $r$-mode oscillations due to
non-linear processes; in this case the quantity
$\tau_\mathrm{Diss}$ includes also dissipation due to these
processes (\citealt{gck13_large}). Using Eq.\ (\ref{tau})
one can show [e.g., by means of equations (16), (19) and
(20) of \citealt{gck13_large}], that the rate of NS
rotation energy ($E_\mathrm{rot}$) loss because of the
instability of $m=2$ $r$-mode is given by
\begin{equation}
\frac{\mathrm d E_\mathrm{rot}}{\mathrm d t}=-3 L_\mathrm{cool}.
\label{dErot_dt}
\end{equation}
Thus, the total NS thermal emissivity $L_\mathrm{cool}$
(in the form of neutrinos from the whole star and thermal
emission from
its atmosphere) constitutes $1/3$ of the rate
of rotation energy loss due to $r$-mode instability.
It is easy to estimate
that, during its life, a ``hot widow''/HOFNAR emits huge
amount of thermal energy,
$E_\mathrm{th}\sim E_\mathrm{rot}/3 \sim 10^{51}$~erg,
while the life time of an NS at this stage
constitutes $\sim E_\mathrm{rot}/(3L_\mathrm{cool})\sim 10^9$~years.%
\footnote{ Accurate evaluation of the time required to
brake the star from rotation frequency 600~Hz down to the
frequency when it becomes stable gives: $\sim 10^9$,
$4\times 10^9$, and $8\times 10^{10}$~yrs for models (a),
(b), and (c), respectively.}
After an NS enters the stability region
oscillation modes rapidly damp out and the star cools down
along the horizontal part of the evolutionary path.

Notice that in the model (a), depending on the value of
$T^\infty_\mathrm{eq}$, NS evolution at the LMXB stage can
be associated with different stability peaks resulting from
the interaction of $r$-mode with different superfluid modes
(\citealt{gck13_large,gck14_short}). For instance, in the
case shown in Fig.\ \ref{Fig_Evol}(a) evolutionary track
for an NS in LMXB will follow the left (low-temperature)
stability peak centered at $T^\infty\approx 4.5\times
10^7$~K if $T^\infty_\mathrm{eq}\lesssim 4\times10^7$~K.
Model (a) explains in this way colder sources, such as IGR
J00291-5934.
If accretion ceases in such system, the produced ``hot
widow''/HOFNAR will descend the low-temperature stability
peak being not so hot.

The analysis presented in this section allow us to conclude that
if some of the observed NSs in LMXBs
are indeed $r$-mode unstable,
then
the new class of objects -- ``hot widows''/HOFNARs --
should emerge
along with MSPs.
This conclusion is
almost insensitive to the actual model
employed to describe the NS instability
and is valid for the models (a)--(c) analyzed here.

\section{``Hot widows''/HOFNARs: observational signatures
} \label{Sec_Nab}

In contrast to MSPs, it can be difficult to measure large
spin frequencies of ``hot widows''/HOFNARs because they
might not display any noticeable pulsed fraction in their
electromagnetic emission.
This is because the resistive relaxation timescale $\tau_r$
for the crustal magnetic field,
\begin{equation}
 \tau_r\approx \frac{4\pi \sigma l^2}{c^2}\sim 4\times 10^6
 \frac{\sigma}{10^{24}\,
 \mathrm{s}^{-1}}\left(\frac{l}{10^5\,
 \mathrm{cm}}\right)\,{\rm yr},
 \label{tau_r}
\end{equation}
is three orders of magnitude lower than the ''hot
widow''/HOFNAR life time due to relatively low electrical
conductivity $\sigma$ at high temperature ($\sigma \sim
10^{24}\, \mathrm{s}^{-1}$ at $T\sim 10^8$~K and density
$\sim 10^{13}\mathrm{\,g\,cm}^{-3}$; see, e.g.,
\citealt*{gyp01,chugunov12}). In Eq.\ (\ref{tau_r}) $c$ is
the speed of light, and $l$ is the typical length scale
$\sim 10^5$~cm.
This simple estimate agrees with much more sophisticated
calculations (e.g., \citealt{uk08,vm09}).
However, one should bear in mind that the problem of
magnetic field relaxation in NSs is a very complicated one,
and
depends on the magnetic field configuration
(which can be nontrivial, see, e.g., \citealt*{pmp11}),
crust conductivity and composition, etc.,
and deserves a special consideration.
Yet, in what follows we assume that the resistive
relaxation leads to decay of the magnetic field in the very
beginning of ``hot widow''/HOFNAR stage making thus
unlikely to measure their frequencies by observing them as
radio pulsars.

On the other hand, ``hot widows''/HOFNARs are hot NSs, with
the internal temperatures $T^\infty\sim10^8$~K
corresponding to the effective redshifted surface
temperatures
$T^\infty_\mathrm{eff}\sim 10^6$~K (\citealt{pcy97}).%
\footnote{This is an additional property that differs them
from MSPs which are also formed as a result of LMXB
evolution, but, as a rule, have lower surface temperatures
(except for the hot spots, see, \citealt{zavlin07,bgr08},
for example). However, one cannot exclude a situation that
a (young) ``hot widow''/HOFNAR can be, at the same time,
classified as MSP. For example, the ``missing link'' binary
PSR J1023+0038 ($\nu = 592.4$~Hz, as reported by
\citealt{Archibald_etal09}) has a relatively high
redshifted surface temperature $T^\infty_\mathrm{eff}\sim
5\times 10^5$~K (\citealt{Homer_etal06,Bogdanov_etal11}).
It is believed that this pulsar is in a binary system,
which leaves the LMXB stage (\citealt{Patruno_etal14}). In
a model (a) it will become a young ``hot widow''/HOFNAR,
attached to the {\it low-temperature} stability peak [see
Fig.\ \ref{Fig_Evol}(a)]. Thus, it can keep its magnetic
field and work as MSP as well.
Another example is the pulsar PSR J1723-2837 ($\nu =
539$~Hz, \citealt{faulkner_etal04}), whose X-ray spectrum
agrees with the redshifted surface temperature
$T^\infty_\mathrm{eff}\sim (4\div 5)\times 10^5$~K (see
table 1 of \citealt{bogdanov_etal14}).
}
Therefore ``hot widows''/HOFNARs should exhibit thermal
X-ray emission from their whole surface%
\footnote{Because there are no reasons for any noticeable
inhomogeneity, the surface temperature of ``hot
widows''/HOFNARs should be uniform, making it difficult to
measure their spin frequencies with X-ray observations.}
and, as they do not accrete matter (and have low magnetic
field), the contribution of the non-thermal component
should be small.
%
%
Exactly these properties of X-ray spectrum have been
required for the selection of candidates to LMXBs in
quiescent state (qLMXBs) among various X-ray sources in GCs
(\citealt{Rutledge_etal00,Heinke_etal_qLMXB03,Guillot_etal11}).%
\footnote{ The only difference is that a nonthermal
contribution of up to $40\%$ is allowed for
qLMXB-candidates (see, e.g.,
\citealt{Heinke_etal_qLMXB03}). However, significant
($\gtrsim 10\%$) nonthermal emission is not necessary for
describing most of the qLMXB candidates (see, e.g., table 2
of \citealt{Heinke_etal_qLMXB03}).}
Thus, qLXMB-candidates, which: ($i$) have never been
observed in outbursts [i.e. no signatures of strong
accretion have been detected; approximately 30 of such
qLMXB candidates are known at the present time, see
\citealt{wdp13}] and ($ii$) have
thermal X-ray spectra,
can be considered as candidates to ``hot widows''/HOFNARs.
It is worth noting, however, that there is a
principal
difference between the ``hot widows''/HOFNARs and real LMXB
systems: ``Hot widow''/HOFNAR's low-mass companion does not
fill the Roche lobe and, as a result, NS does not accrete
significantly. This fact allows us to formulate
identification criteria for ``hot widows''/HOFNARs based on
optical observations (see Sec.\ \ref{Sec_SmokingGun}).

\section{Evidences for the ``hot widows''/HOFNARs existence}
\label{Sec_Evidence}

\subsection{A possibility to reliably
identify ``hot widows''/HOFNARs}
\label{Sec_SmokingGun}

Detection of a hot ($T^\infty_\mathrm{eff}\sim 10^6$~K) NS
in a system, where low-mass companion does not fill the
Roche lobe and hence NS cannot accrete intensively, would
allow one to identify such NS as a ``hot widow''/HOFNAR. In
fact, in that case the only way to explain its high
temperature is to assume that it is permanently heated by
an excited $r$-mode.
Search for such objects is a realistic
task.
For instance, observations show that the source X5 in GC 47
Tuc is a hot NS
($T^\infty_\mathrm{eff}=1.06^{+19}_{-16}\times10^6$~K,
\citealt{hgle03})
in the eclipsing binary with the orbital
period $8.666\pm0.008$~hours (\citealt{hgle02}). Optical
observations with the Hubble Space Telescope by
\citet{Edmonds_etal02_47Tuc_Hubble} have identified the
optical companion as a red main-sequence star and have
revealed the presence of an accretion disk. However, they
also showed, that the companion star considerably
underfills the Roche lobe [according to estimates of
\citet{Edmonds_etal02_47Tuc_Hubble} the ratio of the
companion radius to the size of the Roche lobe is about
$F=0.5-0.6$], and thus is detached from the disk. In that
case accretion can be very small
\citep{Edmonds_etal02_47Tuc_Hubble} and not sufficient to
explain the source's high temperature. Thus, we treat 47
Tuc X5 as a good candidate to  ``hot widows''/HOFNARs.
Another candidate -- the object X7 in GC Tuc 47 -- is a
persistent source with the purely thermal X-ray emission
\citep{heinke_etal05_Chandra_47Tuc,hrng06}. Although
optical observations do not allow to identify its companion
reliably, they also do not contain any indications of the
presence of the accretion disk
\citep{Edmonds_etal02_47Tuc_Hubble}. Moreover, according to
\cite{Edmonds_etal02_47Tuc_Hubble}, if the 5.5 hours period
identified in X-ray observations is real and the companion
underfills its Roche lobe by about the same amount as X5,
then the absolute magnitude of the companion is $M_V\sim
10.6$, and apparent visual magnitude is $V\sim 24.1$, which
is just slightly below the observational upper limit $V\sim
23$.
Thus, this source is also a good candidate to ``hot
widows''/HOFNARs.
Among the other sources we would select five qLMXB
candidates in GCs M28, M13, NGC 5139, NGC 6304 and NGC
6397, studied by \citet{Guillot_etal13},
for which X-ray observations with good signal-to-noise ratio are available.
Except for NGC 6397 U24, all of them can be described with
the purely thermal hydrogen atmosphere spectrum, while for the
source NGC 6397 U24 the contribution of non-thermal
emission in the energy range $0.5-10$~keV is only $\lesssim
2.6^{+1.7}_{-1.7}\%$
\citep{Guillot_etal13}. Moreover, for the source NGC 6397
U24 rather strict constraints can be set on the companion
luminosity in the visible spectral range, $M_V>11$
\citep{Grindlay_etal01}, which indicates that the companion
fills its Roche lobe only if its orbital period is
sufficiently small [e.g., for $M_V\sim 10.6$ and orbital
period 5.5 hours the companion underfills its Roche lobe,
see \citealt{Edmonds_etal02_47Tuc_Hubble}]. Further studies
of the sources mentioned above, as well as other qLMXB
candidates, can allow one to identify some of them as ``hot
widows''/HOFNARs.

\subsection{Indirect evidences} \label{Sec_indirect}

A possibility of existence of ``hot widows''/HOFNARs has
not been previously accounted for in the analysis of
observational data and in population synthesis
calculations. If these objects do exist, they should help
to clarify some inconsistencies between the predictions of
the LMXBs/MSPs evolution theory and observations. We
analyzed the literature on this subject and found a few
such inconsistencies.

Although at least some of those inconsistencies can be
resolved without appeal to the ``hot widow''/HOFNAR
hypothesis we list them here as an {\it indirect evidences}
in its favour.

\begin{itemize}

\item \textbf{ The spin frequencies of accreting NSs in LMXBs
are generally larger
than the frequencies of MSPs},
which are thought to be their descendants.

This fact cannot be explained just by NS spin-down at the
MSP-stage. Possible explanation of this phenomenon was
suggested by \cite{Tauris12} who assumed that the
difference between the spin frequencies of both populations
could be related with
the fast deceleration of an NS in a binary at the moment
when the companion star decouples from its Roche lobe. In
our scenario a complimentary mechanism is possible. It
consists in the formation of ``hot widows''/HOFNARs from
the most rapidly rotating hot NSs in LMXBs. Hence, it
favours the selection of more slowly rotating (stable) NSs
as progenitors of MSPs. A more detailed analysis of MSP
spin distribution by \cite{ptrt14} provides an additional
evidence that this mechanism is relevant. The latter
authors showed that the frequency distribution of the
so-called nuclear-powered MSPs (a subclass of accreting NSs
in LMXBs), whose quasi-coherent oscillations are observed
exclusively during the thermonuclear type-I X-ray bursts
(see, e.g., \citealt{watts12}), differs most strongly from
the frequency distribution of MSPs.
Note that
the most rapidly rotating nuclear-powered MSPs [4U 1608-522, SAX
J1750.8-2900, MXB 1659-298(=X 1658-298), EXO 0748-676, KS
1731-260] are unstable (or marginally stable in the model
of resonance interaction of modes)
with respect to $r$-mode oscillations%
\footnote{The slowest nuclear-powered MSP IGR J17191-2821, is, most
probably, stable (see Fig.\ \ref{Fig_Evol}). The question
of stability is still open for a majority of other
nuclear-powered MSPs (GRS 1741.9-2853, 4U 1636-536, 4U 1728-34, 4U
1702-429) since it is very difficult to extract their
internal temperature from observations. In case of GRS
1741.9-2853 this is because its thermal emission is
strongly absorbed by the high hydrogen column density
\citep{degenaar_et_al_12}; the reason for other sources is
that they are persistent accretors, while we need to
observe them in quiescent state to determine their internal
temperature \citep*{hdh12}.}
and should become ``hot widows''/HOFNARs (rather than MSPs)
in the models (a)--(c) (see Fig.\ \ref{Fig_Evol}). Thus, in
our scenario, spin distribution of nuclear-powered MSPs can
differ significantly from the spin distribution of MSPs, in
agreement with observations. 

\item \textbf{Low-level
non-thermal emission from most of the qLMXB candidates.}

Most of the observed qLMXB candidates do not require a
power-law component for the interpretation of their spectra
\citep{Heinke_etal_qLMXB03,Guillot_etal11}, although the
selection criteria for qLMXBs, applied, e.g., by
\citet{Heinke_etal_qLMXB03}, allows for the contribution of
up to $40\%$ of non-thermal emission. At the same time,
many reliably identified LMXB systems, which have been
observed in outbursts, require a non-thermal component for
the description of their quiescent state (see, e.g., table
2 of \citealt*{dpw12}). A purely thermal spectrum is a
distinguishing feature of ``hot widows''/HOFNARs. That is
why the fact that many qLMXB candidates do have such
spectrum can implicitly indicate that some of them are
``hot widows''/HOFNARs.

\item \textbf{Recurrence time for qLMXB candidates.}

An assumption that all the qLMXB candidates are LMXBs in
quiescence leads to an estimate $N\sim 200$ of the total
number of LMXB systems in GCs \citep{hge05}. However, only
$N_\mathrm{a}=8$ of them have been observed in outburst for
over $t_\mathrm{obs}\sim40$~yrs of satellite observations.
This leads to the following estimate for an average
recurrence time, $\sim N t_\mathrm{obs}/N_\mathrm{a}\,\sim
1000$~yrs \citep{heinke10}.%
\footnote{ At the moment of publication of \citet{heinke10}
only 7 systems have been observed in active state. However,
recently,  the qLMXB candidate X-3 in the GC Ter 5
(classified as Ter 5, W3 in \citet{Heinke_etal_qLMXB03} and
as CX2 in \citet{heinke_etal06_Faint_sources_Ter5}) has
been observed in outburst \citep{Bahramian_etal13},
confirming thus that it is indeed an LMXB system. In this
respect it is interesting to note that this source showed a
variability
\citep{heinke_etal06_Faint_sources_Ter5,Bahramian_etal13}
and its X-ray spectrum had a considerable
[$28^{+8}_{-7}\%$, see table 3 of
\citet{heinke_etal06_Faint_sources_Ter5}] contribution of
power-law component. Thus, it could not be classified as
``hot widow''/HOFNAR candidate even in 2006.}
However, theoretical estimates, based on the disk
instability model, predict much smaller recurrence times
$\lesssim 180$~yrs (see, e.g., section 6.4 in
\citealt{lasota01} and references therein). Possible
resolution of this problem,
related to the low-level
accretion onto NSs, is indicated by \citet{heinke10}.
Still, if some qLMXB candidates are ``hot widows''/HOFNARs,
this reduces an observational estimate for a total number
of LMXB systems in GCs and, as a consequence, makes the
recurrence time smaller.

\item \textbf{Total number of MSPs in GCs.}

\citet{ivanova_etal08} showed that account for all channels
of formation and spin up of NSs in GCc leads to
overproduction of MSPs, although all these channels are
necessary for explanation of the observed number of LMXB
systems. \cite{ivanova_etal08} suggested that the formation
of MSPs with large magnetic fields, leading to their fast
deceleration, could serve as a possible way out of that
paradox.
However, almost all the observed MSPs have low spin-down
rates (see, e.g., \citealt{Tauris12}). An existence of
``hot widows''/HOFNARs can help to resolve the paradox
because of the following two reasons: ($i$) if some of
qLMXB candidates are indeed ``hot widows''/HOFNARs this
should reduce an observational estimate for the total
number of LMXBs in GCs; ($ii$) a possibility that ``hot
widows''/HOFNARs   originate from LMXBs should result in a
lower production rate of MSPs.
\end{itemize}

Detailed population synthesis calculations,
taking into account a possibility of ``hot widow''/HOFNAR formation,
can reveal additional evidences for the existence of these objects.

\section{Results, conclusions and outlook}
\label{Sec_Summ}

We study a hypothesis about a new possible class of hot
rapidly rotating non-accreting neutron stars (``hot
widow''/HOFNAR objects), whose high temperature is
maintained for a long time ($\sim 10^9$~years) by
instability of $r$-modes.
In Sec.\ \ref{Sec_Evol} we show that these objects should
inevitably be produced in LMXBs, provided that the $r$-mode
instability operates for NSs with the spin frequencies
$300\div 600$~Hz (which means that at least some of the
observed NSs are unstable).
The X-ray sources, whose spectra allow to classify them as
qLMXB candidates, can equally well be considered as
candidates for ``hot widows''/HOFNARs (Sec.\
\ref{Sec_Nab}).
Combined analysis of X-ray and optical observations of any
such candidate can exclude the possibility of Roche-lobe
overflow in a binary system and hence prove that this NS is
indeed a ``hot widow''/HOFNAR.
Even observations, available at the present time, indicate
that this condition is met for the sources X5 and X7 in GC
47 Tuc \citep{Edmonds_etal02_47Tuc_Hubble}.
This makes them promising candidates for ``hot
widows''/HOFNARs. In addition, a statistical analysis of
observational data and the results of population synthesis
calculations provide indirect evidences in favour of the
``hot widow''/HOFNAR existence (Sec.\ \ref{Sec_indirect}).

A confirmation of the ``hot widow''/HOFNAR hypothesis opens
interesting new ways for studying NSs. If these objects are
real,
this would provide a strong evidence for the 
CFS-instability in NSs and hence the possibility of
gravitational wave generation by {\it mass-current}
multipole (rather than by \textit{mass} multipole, as is
usually the case, see, e.g., \citealt{Andersson_etal13}).
Moreover, ``hot widows''/HOFNARs have stable purely thermal
emission spectra and thus can be used, together with qLMXBs
\citep{Guillot_etal13}, to constrain mass and radius of NSs
from their X-ray observations. Furthermore, in the model of
resonance interaction of oscillation modes [Fig.\
\ref{Fig_Evol}(a) and \citealt{gck14_short}] ``hot
widow''/HOFNAR temperatures should coincide with the
resonance temperatures in oscillation spectrum of rotating
NSs. The resonance temperatures are sensitive to the
equation of state, critical temperature profiles etc.
Therefore, confronting theoretically calculated resonance
temperatures with observations can serve as a unique
seismological tool to study microphysical properties of NS
interiors. The fact that the very same objects allow, in
principle, to measure their masses and radii with X-ray
spectra \citep{Guillot_etal13} makes this method even more
promising.


 At the present time, all known ``hot widow''/HOFNAR
candidates are localized in GCs. The situation can change
with SRG X-ray observatory, which will perform all-sky
survey with high spatial and spectral resolution. It will
make possible to identify hot NSs (see, e.g.,
\citealt{SRG}), which can be considered as qLMXB
candidates, ``hot widows''/HOFNARs, or young NSs. Detailed
analysis of these candidates would be a very interesting
task. It could help  to solve the MSP birthrate problem
\citep{kn88,Levin_etal13}, and present one more evidence
for the ``hot widow''/HOFNAR existence.

\section*{Acknowledgements}
We are grateful to
A.~A.~Danilenko,
C.~O. Heinke,
O.~Y.~Kargaltsev,
G.~G.~Pavlov,
Y.~A.~Shibanov,
D.~G.~Yakovlev,
and D.~A.~Zyuzin for
insightful comments and discussions,
and to O.~V. Zakutnyaya for assistance
in preparation of the manuscript.
This work was
supported by RSCF grant 14-12-00316.

\bibliographystyle{mn2e}

\begin{thebibliography}{76}
\expandafter\ifx\csname
natexlab\endcsname\relax\def\natexlab#1{#1}\fi

\bibitem[{{Andersson}(1998)}]{andersson98}
{Andersson} N., 1998, \apj, 502, 708

\bibitem[{{Andersson} {et~al}\mbox{.}(2013){Andersson}, {Baker}, {Belczynski},
  {Bernuzzi}, {Berti}, {Cadonati}, {Cerd{\'a}-Dur{\'a}n}, {Clark}, {Favata},
  {Finn}, {Fryer}, {Giacomazzo}, {Gonz{\'a}lez}, {Hendry}, {Heng}, {Hild},
  {Johnson-McDaniel}, {Kalmus}, {Klimenko}, {Kobayashi}, {Kokkotas}, {Laguna},
  {Lehner}, {Levin}, {Liebling}, {MacFadyen}, {Mandel}, {Marka}, {Marka},
  {Neilsen}, {O'Brien}, {Perna}, {Read}, {Reisswig}, {Rodriguez}, {Ruffert},
  {Schnetter}, {Searle}, {Shawhan}, {Shoemaker}, {Soderberg}, {Sperhake},
  {Sutton}, {Tanvir}, {Was}, \& {Whitcomb}}]{Andersson_etal13}
{Andersson} N. {et~al.}, 2013, Classical and Quantum
Gravity, 30, 193002

\bibitem[{{Andersson} \& {Kokkotas}(2001)}]{ak01}
{Andersson} N., {Kokkotas} K.~D., 2001, International
Journal of Modern Physics
  D, 10, 381

\bibitem[{{Antoniadis} {et~al}\mbox{.}(2013){Antoniadis}, {Freire}, {Wex},
  {Tauris}, {Lynch}, {van Kerkwijk}, {Kramer}, {Bassa}, {Dhillon}, {Driebe},
  {Hessels}, {Kaspi}, {Kondratiev}, {Langer}, {Marsh}, {McLaughlin},
  {Pennucci}, {Ransom}, {Stairs}, {van Leeuwen}, {Verbiest}, \&
  {Whelan}}]{Antoniadis_etal13}
{Antoniadis} J. {et~al.}, 2013, Science, 340, 448

\bibitem[{{Archibald} {et~al}\mbox{.}(2009){Archibald}, {Stairs}, {Ransom},
  {Kaspi}, {Kondratiev}, {Lorimer}, {McLaughlin}, {Boyles}, {Hessels}, {Lynch},
  {van Leeuwen}, {Roberts}, {Jenet}, {Champion}, {Rosen}, {Barlow}, {Dunlap},
  \& {Remillard}}]{Archibald_etal09}
{Archibald} A.~M. {et~al.}, 2009, Science, 324, 1411

\bibitem[{{Bahramian} {et~al}\mbox{.}(2014){Bahramian}, {Heinke}, {Sivakoff},
  {Altamirano}, {Wijnands}, {Homan}, {Linares}, {Pooley}, {Degenaar}, \&
  {Gladstone}}]{Bahramian_etal13}
{Bahramian} A. {et~al.}, 2014, \apj, 780, 127

\bibitem[{{Bogdanov} {et~al}\mbox{.}(2011){Bogdanov}, {Archibald}, {Hessels},
  {Kaspi}, {Lorimer}, {McLaughlin}, {Ransom}, \& {Stairs}}]{Bogdanov_etal11}
{Bogdanov} S., {Archibald} A.~M., {Hessels} J.~W.~T.,
{Kaspi} V.~M., {Lorimer}
  D., {McLaughlin} M.~A., {Ransom} S.~M., {Stairs} I.~H., 2011, \apj, 742, 97

\bibitem[{{Bogdanov} {et~al}\mbox{.}(2014){Bogdanov}, {Esposito}, {Crawford},
  {Possenti}, {McLaughlin}, \& {Freire}}]{bogdanov_etal14}
{Bogdanov} S., {Esposito} P., {Crawford}, III F.,
{Possenti} A., {McLaughlin}
  M.~A., {Freire} P., 2014, \apj, 781, 6

\bibitem[{{Bogdanov} {et~al}\mbox{.}(2008){Bogdanov}, {Grindlay}, \&
  {Rybicki}}]{bgr08}
{Bogdanov} S., {Grindlay} J.~E., {Rybicki} G.~B., 2008,
\apj, 689, 407

\bibitem[{{Bondarescu} \& {Wasserman}(2013)}]{bw13}
{Bondarescu} R., {Wasserman} I., 2013, \apj, 778, 9

\bibitem[{{Brown} {et~al}\mbox{.}(1998){Brown}, {Bildsten}, \&
  {Rutledge}}]{bbr98}
{Brown} E.~F., {Bildsten} L., {Rutledge} R.~E., 1998,
\apjl, 504, L95

\bibitem[{{Chandrasekhar}(1970)}]{chandrasekhar70a}
{Chandrasekhar} S., 1970, \prl, 24, 611

\bibitem[{{Chen} {et~al}\mbox{.}(2013){Chen}, {Chen}, {Tauris}, \&
  {Han}}]{ccth13}
{Chen} H.-L., {Chen} X., {Tauris} T.~M., {Han} Z., 2013,
\apj, 775, 27

\bibitem[\protect\citeauthoryear{Chugunov}{2012}]{chugunov12}
Chugunov A.I.,  2012, Astron.\
 Lett., 38, 25

\bibitem[{{Degenaar} {et~al}\mbox{.}(2012{\natexlab{a}}){Degenaar}, {Patruno},
  \& {Wijnands}}]{dpw12}
{Degenaar} N., {Patruno} A., {Wijnands} R.,
2012{\natexlab{a}}, \apj, 756, 148

\bibitem[{{Degenaar} {et~al}\mbox{.}(2012{\natexlab{b}}){Degenaar}, {Wijnands},
  {Cackett}, {Homan}, {in't Zand}, {Kuulkers}, {Maccarone}, \& {van der
  Klis}}]{degenaar_et_al_12}
{Degenaar} N., {Wijnands} R., {Cackett} E.~M., {Homan} J.,
{in't Zand}
  J.~J.~M., {Kuulkers} E., {Maccarone} T.~J., {van der Klis} M.,
  2012{\natexlab{b}}, \aap, 545, A49

\bibitem[{{Demorest} {et~al}\mbox{.}(2010){Demorest}, {Pennucci}, {Ransom},
  {Roberts}, \& {Hessels}}]{Demorest_etal10}
{Demorest} P.~B., {Pennucci} T., {Ransom} S.~M., {Roberts}
M.~S.~E., {Hessels}
  J.~W.~T., 2010, \nat, 467, 1081

\bibitem[{{Edmonds} {et~al}\mbox{.}(2002){Edmonds}, {Heinke}, {Grindlay}, \&
  {Gilliland}}]{Edmonds_etal02_47Tuc_Hubble}
{Edmonds} P.~D., {Heinke} C.~O., {Grindlay} J.~E.,
{Gilliland} R.~L., 2002,
  \apjl, 564, L17

\bibitem[{{Faulkner} {et~al}\mbox{.}(2004){Faulkner}, {Stairs}, {Kramer},
  {Lyne}, {Hobbs}, {Possenti}, {Lorimer}, {Manchester}, {McLaughlin},
  {D'Amico}, {Camilo}, \& {Burgay}}]{faulkner_etal04}
{Faulkner} A.~J. {et~al.}, 2004, \mnras, 355, 147

\bibitem[{{Friedman} \& {Morsink}(1998)}]{fm98}
{Friedman} J.~L., {Morsink} S.~M., 1998, \apj, 502, 714

\bibitem[{{Friedman} \& {Schutz}(1978{\natexlab{a}})}]{fs78a}
{Friedman} J.~L., {Schutz} B.~F., 1978{\natexlab{a}}, \apj,
221, 937

\bibitem[{{Friedman} \& {Schutz}(1978{\natexlab{b}})}]{fs78b}
{Friedman} J.~L., {Schutz} B.~F., 1978{\natexlab{b}}, \apj,
222, 281

\bibitem[\protect\citeauthoryear{Gnedin, Yakovlev \& Potekhin}
{Gnedin et al.}{2001}]{gyp01}
    Gnedin O.Y., Yakovlev D.G.,
    Potekhin A.Y., 2001, MNRAS, 324, 725


\bibitem[{{Grindlay} {et~al}\mbox{.}(2001){Grindlay}, {Heinke}, {Edmonds},
  {Murray}, \& {Cool}}]{Grindlay_etal01}
{Grindlay} J.~E., {Heinke} C.~O., {Edmonds} P.~D., {Murray}
S.~S., {Cool}
  A.~M., 2001, \apjl, 563, L53

\bibitem[{{Guillot} {et~al}\mbox{.}(2011){Guillot}, {Rutledge}, {Brown},
  {Pavlov}, \& {Zavlin}}]{Guillot_etal11}
{Guillot} S., {Rutledge} R.~E., {Brown} E.~F., {Pavlov}
G.~G., {Zavlin} V.~E.,
  2011, \apj, 738, 129

\bibitem[{{Guillot} {et~al}\mbox{.}(2013){Guillot}, {Servillat}, {Webb}, \&
  {Rutledge}}]{Guillot_etal13}
{Guillot} S., {Servillat} M., {Webb} N.~A., {Rutledge}
R.~E., 2013, \apj, 772,
  7


\bibitem[{{Gusakov} {et~al}\mbox{.}(2014{\natexlab{a}}){Gusakov}, {Chugunov}, \&
  {Kantor}}]{gck14_short}
{Gusakov} M.~E., {Chugunov} A.~I., {Kantor} E.~M.,
2014{\natexlab{a}}, \prl, 112, 151101


\bibitem[{{Gusakov} {et~al}\mbox{.}(2014{\natexlab{b}}){Gusakov}, {Chugunov}, \&
  {Kantor}}]{gck13_large}
{Gusakov} M.~E., {Chugunov} A.~I., {Kantor} E.~M.,
2014{\natexlab{b}}, \prd, accepted for publication,
arXiv:1305.3825

\bibitem[{{Gusakov} {et~al}\mbox{.}(2004){Gusakov}, {Kaminker}, {Yakovlev}, \&
  {Gnedin}}]{gkyg04}
{Gusakov} M.~E., {Kaminker} A.~D., {Yakovlev} D.~G.,
{Gnedin} O.~Y., 2004,
  \aap, 423, 1063

\bibitem[{Haensel {et~al}\mbox{.}(2006)Haensel, Potekhin, \& Yakovlev}]{hpy07}
Haensel P., Potekhin A., Yakovlev D., 2006, Neutron Stars
1: Equation of State
  and Structure, Astrophysics and Space Science Library. Springer

\bibitem[{{Harding}(2013)}]{Harding13}
{Harding} A.~K., 2013, Frontiers of Physics, 8, 679

\bibitem[{{Haskell} {et~al}\mbox{.}(2012){Haskell}, {Degenaar}, \&
  {Ho}}]{hdh12}
{Haskell} B., {Degenaar} N., {Ho} W.~C.~G., 2012, \mnras,
424, 93

\bibitem[{{Haskell} {et~al}\mbox{.}(2014){Haskell}, {Glampedakis}, \&
  {Andersson}}]{hga14}
Haskell B., Glampedakis K., Andersson N., 2014, \mnras,
441, 1662

\bibitem[{{Heinke}(2010)}]{heinke10}
{Heinke} C.~O., 2010, in American Institute of Physics
Conference Series, Vol.
  1314, American Institute of Physics Conference Series, {Kologera} V., {van
  der Sluys} M., eds., pp. 135--142

\bibitem[{{Heinke} {et~al}\mbox{.}(2005{\natexlab{a}}){Heinke}, {Grindlay}, \&
  {Edmonds}}]{hge05}
{Heinke} C.~O., {Grindlay} J.~E., {Edmonds} P.~D.,
2005{\natexlab{a}}, \apj,
  622, 556

\bibitem[{{Heinke} {et~al}\mbox{.}(2005{\natexlab{b}}){Heinke}, {Grindlay},
  {Edmonds}, {Cohn}, {Lugger}, {Camilo}, {Bogdanov}, \&
  {Freire}}]{heinke_etal05_Chandra_47Tuc}
{Heinke} C.~O., {Grindlay} J.~E., {Edmonds} P.~D., {Cohn}
H.~N., {Lugger}
  P.~M., {Camilo} F., {Bogdanov} S., {Freire} P.~C., 2005{\natexlab{b}}, \apj,
  625, 796

\bibitem[{{Heinke} {et~al}\mbox{.}(2002){Heinke}, {Grindlay}, {Lloyd}, \&
  {Edmonds}}]{hgle02}
{Heinke} C.~O., {Grindlay} J.~E., {Lloyd} D.~A., {Edmonds}
P.~D., 2002, in
  Astronomical Society of the Pacific Conference Series, Vol. 271, Neutron
  Stars in Supernova Remnants, {Slane} P.~O., {Gaensler} B.~M., eds., p. 349

\bibitem[{{Heinke} {et~al}\mbox{.}(2003{\natexlab{a}}){Heinke}, {Grindlay},
  {Lloyd}, \& {Edmonds}}]{hgle03}
{Heinke} C.~O., {Grindlay} J.~E., {Lloyd} D.~A., {Edmonds}
P.~D.,
  2003{\natexlab{a}}, \apj, 588, 452

\bibitem[{{Heinke} {et~al}\mbox{.}(2003{\natexlab{b}}){Heinke}, {Grindlay},
  {Lugger}, {Cohn}, {Edmonds}, {Lloyd}, \& {Cool}}]{Heinke_etal_qLMXB03}
{Heinke} C.~O., {Grindlay} J.~E., {Lugger} P.~M., {Cohn}
H.~N., {Edmonds}
  P.~D., {Lloyd} D.~A., {Cool} A.~M., 2003{\natexlab{b}}, \apj, 598, 501

\bibitem[{{Heinke} \& {Ho}(2010)}]{hh10}
{Heinke} C.~O., {Ho} W.~C.~G., 2010, Astrophys. J. Lett.,
719, L167

\bibitem[{{Heinke} {et~al}\mbox{.}(2009){Heinke}, {Jonker}, {Wijnands},
  {Deloye}, \& {Taam}}]{heinke_et_al_09}
{Heinke} C.~O., {Jonker} P.~G., {Wijnands} R., {Deloye}
C.~J., {Taam} R.~E.,
  2009, \apj, 691, 1035

\bibitem[{{Heinke} {et~al}\mbox{.}(2007){Heinke}, {Jonker}, {Wijnands}, \&
  {Taam}}]{hjwt07}
{Heinke} C.~O., {Jonker} P.~G., {Wijnands} R., {Taam}
R.~E., 2007, \apj, 660,
  1424

\bibitem[{{Heinke} {et~al}\mbox{.}(2006{\natexlab{a}}){Heinke}, {Rybicki},
  {Narayan}, \& {Grindlay}}]{hrng06}
{Heinke} C.~O., {Rybicki} G.~B., {Narayan} R., {Grindlay}
J.~E.,
  2006{\natexlab{a}}, \apj, 644, 1090

\bibitem[{{Heinke} {et~al}\mbox{.}(2006{\natexlab{b}}){Heinke}, {Wijnands},
  {Cohn}, {Lugger}, {Grindlay}, {Pooley}, \&
  {Lewin}}]{heinke_etal06_Faint_sources_Ter5}
{Heinke} C.~O., {Wijnands} R., {Cohn} H.~N., {Lugger}
P.~M., {Grindlay} J.~E.,
  {Pooley} D., {Lewin} W.~H.~G., 2006{\natexlab{b}}, \apj, 651, 1098

\bibitem[{{Heyl}(2002)}]{heyl02}
{Heyl} J.~S., 2002, \apjl, 574, L57

\bibitem[{{Ho} \& {Andersson}(2012)}]{ha12}
{Ho} W.~C.~G., {Andersson} N., 2012, Nature Physics, 8, 787

\bibitem[{{Ho} {et~al}\mbox{.}(2011) {Ho}, {Andersson}, \& {Haskell}}]{hah11}
{Ho} W.~C.~G., {Andersson} N., Haskell B, 2011, \prl, 107,
101101




\bibitem[{{Ho} \& {Lai}(2000)}]{hl00}
{Ho} W.~C.~G., {Lai} D., 2000, \apj, 543, 386

\bibitem[{{Homer} {et~al}\mbox{.}(2006){Homer}, {Szkody}, {Chen}, {Henden},
  {Schmidt}, {Anderson}, {Silvestri}, \& {Brinkmann}}]{Homer_etal06}
{Homer} L., {Szkody} P., {Chen} B., {Henden} A., {Schmidt}
G., {Anderson}
  S.~F., {Silvestri} N.~M., {Brinkmann} J., 2006, \aj, 131, 562

\bibitem[{{Illarionov} \& {Sunyaev}(1975)}]{is75}
{Illarionov} A.~F., {Sunyaev} R.~A., 1975, \aap, 39, 185

\bibitem[{{Ivanova} {et~al}\mbox{.}(2008){Ivanova}, {Heinke}, {Rasio},
  {Belczynski}, \& {Fregeau}}]{ivanova_etal08}
{Ivanova} N., {Heinke} C.~O., {Rasio} F.~A., {Belczynski}
K., {Fregeau} J.~M.,
  2008, \mnras, 386, 553

\bibitem[{{Kantor}(2011)}]{Kantor11}
{Kantor} E.~M., 2011, in International Conference on the
PHYSICS OF NEUTRON
  STARS: Book of Abstracts, {Baiko} D.~A., {Uvarov} Y.~A., {Yakovlev} D.~G.,
  eds., p.\ 64.

\bibitem[{{Kulkarni} \& {Narayan}(1988)}]{kn88}
{Kulkarni} S.~R., {Narayan} R., 1988, \apj, 335, 755

\bibitem[{{Lasota}(2001)}]{lasota01}
{Lasota} J.-P., 2001, \nar, 45, 449

\bibitem[{{Levin} {et~al}\mbox{.}(2013){Levin}, {Bailes}, {Barsdell}, {Bates},
  {Bhat}, {Burgay}, {Burke-Spolaor}, {Champion}, {Coster}, {D'Amico},
  {Jameson}, {Johnston}, {Keith}, {Kramer}, {Milia}, {Ng}, {Possenti},
  {Stappers}, {Thornton}, \& {van Straten}}]{Levin_etal13}
{Levin} L. {et~al.}, 2013, \mnras, 434, 1387

\bibitem[{{Levin}(1999)}]{levin99}
{Levin} Y., 1999, \apj, 517, 328

\bibitem[{{Levin} \& {Ushomirsky}(2001)}]{lu01}
{Levin} Y., {Ushomirsky} G., 2001, \mnras, 324, 917

\bibitem[{{Lindblom} {et~al}\mbox{.}(1998){Lindblom}, {Owen}, \&
  {Morsink}}]{lom98}
{Lindblom} L., {Owen} B.~J., {Morsink} S.~M., 1998, \prl,
80, 4843

\bibitem[{{Mahmoodifar} \& {Strohmayer}(2013)}]{ms13}
{Mahmoodifar} S., {Strohmayer} T., 2013, \apj, 773, 140

\bibitem[{{Merloni} {et~al}\mbox{.}(2012){Merloni}, {Predehl}, {Becker},
  {B{\"o}hringer}, {Boller}, {Brunner}, {Brusa}, {Dennerl}, {Freyberg},
  {Friedrich}, {Georgakakis}, {Haberl}, {Hasinger}, {Meidinger}, {Mohr},
  {Nandra}, {Rau}, {Reiprich}, {Robrade}, {Salvato}, {Santangelo}, {Sasaki},
  {Schwope}, {Wilms}, \& {German eROSITA Consortium}}]{SRG}
{Merloni} A. {et~al.}, 2012, ArXiv:1209.3114

\bibitem[{{Owen} {et~al}\mbox{.}(1998){Owen}, {Lindblom}, {Cutler}, {Schutz},
  {Vecchio}, \& {Andersson}}]{olcsva98}
{Owen} B.~J., {Lindblom} L., {Cutler} C., {Schutz} B.~F.,
{Vecchio} A.,
  {Andersson} N., 1998, \prd, 58, 084020

\bibitem[{{Page} {et~al}\mbox{.}(2004){Page}, {Lattimer}, {Prakash}, \&
  {Steiner}}]{page04}
{Page} D., {Lattimer} J.~M., {Prakash} M., {Steiner} A.~W.,
2004, \apjs, 155,
  623

\bibitem[{{Page} {et~al}\mbox{.}(2011){Page}, {Prakash}, {Lattimer}, \&
  {Steiner}}]{page11}
{Page} D., {Prakash} M., {Lattimer} J.~M., {Steiner} A.~W.,
2011, \prl, 106,
  081101

\bibitem[{{Papitto} {et~al}\mbox{.}(2013){Papitto}, {Ferrigno}, {Bozzo}, {Rea},
  {Pavan}, {Burderi}, {Burgay}, {Campana}, {di Salvo}, {Falanga},
  {Filipovi{\'c}}, {Freire}, {Hessels}, {Possenti}, {Ransom}, {Riggio},
  {Romano}, {Sarkissian}, {Stairs}, {Stella}, {Torres}, {Wieringa}, \&
  {Wong}}]{Papitto_etal13}
{Papitto} A. {et~al.}, 2013, \nat, 501, 517

\bibitem[{{Papitto} {et~al}\mbox{.}(2014){Papitto}, {Torres}, {Rea}, \&
  {Tauris}}]{ptrt14}
{Papitto} A., {Torres} D.~F., {Rea} N., {Tauris} T., 2014,
ArXiv:1403.6775

\bibitem[{{Patruno} {et~al}\mbox{.}(2014){Patruno}, {Archibald}, {Hessels},
  {Bogdanov}, {Stappers}, {Bassa}, {Janssen}, {Kaspi}, {Tendulkar}, \&
  {Lyne}}]{Patruno_etal14}
{Patruno} A. {et~al.}, 2014, \apjl, 781, L3

\bibitem[{{Pines} \& {Alpar}(1985)}]{pa85}
{Pines} D., {Alpar} M.~A., 1985, \nat, 316, 27

\bibitem[{{Posselt} {et~al}\mbox{.}(2013){Posselt}, {Pavlov}, {Suleimanov}, \&
  {Kargaltsev}}]{ppsk13}
{Posselt} B., {Pavlov} G.~G., {Suleimanov} V., {Kargaltsev}
O., 2013, \apj,
  779, 186

\bibitem[{{Potekhin} {et~al}\mbox{.}(1997){Potekhin}, {Chabrier}, \&
  {Yakovlev}}]{pcy97}
{Potekhin} A.~Y., {Chabrier} G., {Yakovlev} D.~G., 1997,
\aap, 323, 415


\bibitem[{{Priymak} {et~al}\mbox{.}(2011){Priymak}, {Melatos}, \&
{Payne}}]{pmp11}
    Priymak M., Melatos A., Payne, D. J. B., 2011,
    \mnras, 417, 2696

\bibitem[{{Reisenegger} \& {Bona{\v c}i{\'c}}(2003)}]{rb03}
{Reisenegger} A., {Bona{\v c}i{\'c}} A., 2003, \prl, 91,
201103

\bibitem[{{Rutledge} {et~al}\mbox{.}(2000){Rutledge}, {Bildsten}, {Brown},
  {Pavlov}, \& {Zavlin}}]{Rutledge_etal00}
{Rutledge} R.~E., {Bildsten} L., {Brown} E.~F., {Pavlov}
G.~G., {Zavlin} V.~E.,
  2000, \apj, 529, 985

\bibitem[{{Shternin} {et~al}\mbox{.}(2011){Shternin}, {Yakovlev}, {Heinke},
  {Ho}, \& {Patnaude}}]{shternin11}
{Shternin} P.~S., {Yakovlev} D.~G., {Heinke} C.~O., {Ho}
W.~C.~G., {Patnaude}
  D.~J., 2011, \mnras, 412, L108

\bibitem[{{Tauris}(2011)}]{Tauris11}
{Tauris} T.~M., 2011, in Astronomical Society of the
Pacific Conference Series,
  Vol. 447, Evolution of Compact Binaries, {Schmidtobreick} L., {Schreiber}
  M.~R., {Tappert} C., eds., p. 285

\bibitem[{{Tauris}(2012)}]{Tauris12}
{Tauris} T.~M., 2012, Science, 335, 561

\bibitem[{{Tauris} \& {van den Heuvel}(2006)}]{tvh06}
{Tauris} T.~M., {van den Heuvel} E.~P.~J., 2006, {Formation
and evolution of
  compact stellar X-ray sources}, {Lewin} W.~H.~G., {van der Klis} M., eds.,
  pp. 623--665

\bibitem[{{Urpin} \& {Konenkov}(2008)}]{uk08}
{Urpin} V., {Konenkov} D., 2008, \aap, 483, 223

\bibitem[{{Vigan{\`o}} {et~al}\mbox{.}(2013){Vigan{\`o}}, {Rea}, {Pons},
  {Perna}, {Aguilera}, \& {Miralles}}]{Vigano_etal13_Unif}
{Vigan{\`o}} D., {Rea} N., {Pons} J.~A., {Perna} R.,
{Aguilera} D.~N.,
  {Miralles} J.~A., 2013, \mnras, 434, 123

\bibitem[{{Vigelius \& Melatos}(2009)}]{vm09} Vigelius M., Melatos A., 2009,
\mnras 395, 1985


\bibitem[{{Watts}(2012)}]{watts12}
{Watts} A.~L., 2012, \araa, 50, 609

\bibitem[{{Wijnands} {et~al}\mbox{.}(2013){Wijnands}, {Degenaar}, \&
  {Page}}]{wdp13}
{Wijnands} R., {Degenaar} N., {Page} D., 2013, \mnras, 432,
2366

\bibitem[{{Zavlin}(2007)}]{zavlin07}
{Zavlin} V.~E., 2007, \apss, 308, 297

\end{thebibliography}

\label{lastpage}
\end{document}